\documentclass[10pt,english,aps,amssymb,prb,twocolumn,showpacs]{revtex4-1}
\pdfoutput=1
\usepackage[utf8]{inputenc}
\usepackage{amsmath}
\usepackage{graphicx}
\usepackage{amssymb}
\usepackage{esint}
\usepackage{verbatim}
\usepackage{bm}
\usepackage{bbm}

\usepackage{color}

\makeatletter
\@ifundefined{textcolor}{}
{
 \definecolor{WHITE}{gray}{1}
 \definecolor{RED}{rgb}{1,0,0}
 \definecolor{GREEN}{rgb}{0,1,0}
 \definecolor{BLUE}{rgb}{0,0,1}
 \definecolor{CYAN}{cmyk}{1,0,0,0}
 \definecolor{MAGENTA}{cmyk}{0,1,0,0}
 \definecolor{YELLOW}{cmyk}{0,0,1,0}
 }

\DeclareMathOperator{\IM}{Im}
\DeclareMathOperator{\RE}{Re}
\DeclareMathOperator{\sign}{\text{sgn}}
\renewcommand{\phi}{\varphi}
\renewcommand{\epsilon}{\varepsilon}

\renewcommand{\vec}[1]{{\bf #1}}

\newcommand{\mc}{\mathcal}

\usepackage{babel}

\begin{document}
\title {Engineering 1D topological phases on $p$-wave superconductors}
\author{Isac Sahlberg}
\author{Alex Westström}
\author{Kim Pöyhönen}
\author{Teemu Ojanen}
\email[Correspondence to ]{teemuo@boojum.hut.fi}
\affiliation{Department of Applied Physics (LTL), Aalto University, P.~O.~Box 15100,
FI-00076 AALTO, Finland }
\date{\today}
\begin{abstract}
In this work, we study how, with the aid of impurity engineering, two-dimensional $p$-wave superconductors can be employed as a platform for one-dimensional topological phases. We discover that, while chiral and helical parent states themselves are topologically nontrivial, a chain of scalar impurities on both systems support multiple topological phases and Majorana end states. We develop an approach which allows us to extract the topological invariants and subgap spectrum, even away from the center of the gap, for the representative cases of spinless, chiral and helical superconductors. We find that the magnitude of the topological gaps protecting the nontrivial phases may be a significant fraction of the gap of the underlying superconductor. 
       
\end{abstract}
\pacs{73.63.Nm,74.50.+r,74.78.Na,74.78.Fk}
\maketitle
\bigskip{}

\section{introduction}

The contemporary search for novel phases of matter is beginning to employ a designer-matter approach. According to the new paradigm, the restrictions of available materials in Nature no longer pose fundamental obstacles in realizing new quantum phases of matter and emergent particles.  By fabricating structures combining different materials and geometries it is possible to access systems that are limited only by our imagination and ability to manipulate materials.  The efforts to realize topological superconductors is a prime example of engineering of novel quantum phases of matter.\cite{oreg:2010:1,lutchyn:2010:1,mourik:2012:1,das:2012:1}

Recent efforts in designing topological superconductors have identified arrays of magnetic atoms on $s$-wave superconducting surfaces as a promising platform for one-dimensional (1d) and two-dimensional (2d) topological superconductivity. This mechanism is based on the fact that isolated magnetic atoms induce Yu-Shiba-Rusinov states in the superconducting gap.\cite{yu:1965:1,shiba:1968:1,rusinov:1969:1,balatsky:2006:1,yazdani:1997:1} A chain of magnetic moments form subgap bands that may undergo topological phase transitions.\cite{choy:2011:1,nadj-perge:2013:1,pientka:2013:1,poyhonen:2014:1,vazifeh:2013:1,klinovaja:2013:1,braunecker:2013:1,heimes:2014:1,brydon:2015:1,nadj-perge:2014:1,pawlak:2016:1} The necessary ingredients for non-trivial topology are an appropriate magnetic texture of the chain and, in the case of ferromagnetic chains, a spin-orbit coupling on the surface.  

The need for magnetic structures arises from the well-known robustness of conventional $s$-wave superconductors to non-magnetic impurities. While examples of conventional superconductors are ample, the need for robust magnetic textures might prove to be an obstacle. This can be circumvented by employing an unconventional superconductor where scalar impurities induce subgap states. This has motivated the recent interest in topological state engineering in $p$-wave superconductors by potential impurities.\cite{kimme:2016:1,neupert:2016:1,kaladzhyan:2016:1} 

While various 2d $p$-wave superconductors are intrinsically topologically nontrivial,\cite{volovik, bernevig} a suitable patterning with scalar impurities may lead to a zoo of nontrivial descendants.\cite{kimme:2016:1,kaladzhyan:2016:1} 
Also, as discussed below, 1d topological subsystems and networks could offer better access to exotic Majorana physics than the parent 2d topological phase, not least due to the possibility for higher energy gaps. In this work we study $p$-wave superconductors decorated by point-like potential impurities arranged in a linear chain. In contrast to previous works, we will derive a theoretical framework which allows us to calculate the spectrum of a full microscopic long-range hopping model for all subgap energies. We solve the topological phase diagram and discover three nontrivial 1d phases distinguished by the number of Majorana bound states (MBS) in finite chains. The main result of the paper is that the topological gaps separating the MBS from other states may be a significant fraction of the gap of the underlying superconductor. This is in a stark contrast to the well-known MBS at the vortex cores of chiral $p$-wave superconductors which are separated from the other states by a minigap which is tiny in typical circumstances where the pairing gap is small compared to the Fermi energy.\cite{kopnin:1991:1} 

In Sec.~II  we formulate the problem of an impurity chain on a spinless chiral $p$-wave superconductor and develop the mathematical tools to solve the spectrum for all subgap energies.  In Sec.~III we calculate the topological phase diagrams and identify three different phases distinguished by the number of MBS. The results for spinless fermions can be readily generalized to chiral and helical superconductors with spin, as explained in Sec.~IV.  Our findings are discussed and summarized in Sec.~\ref{summ}. 

\begin{figure}
\includegraphics[width=0.8\linewidth]{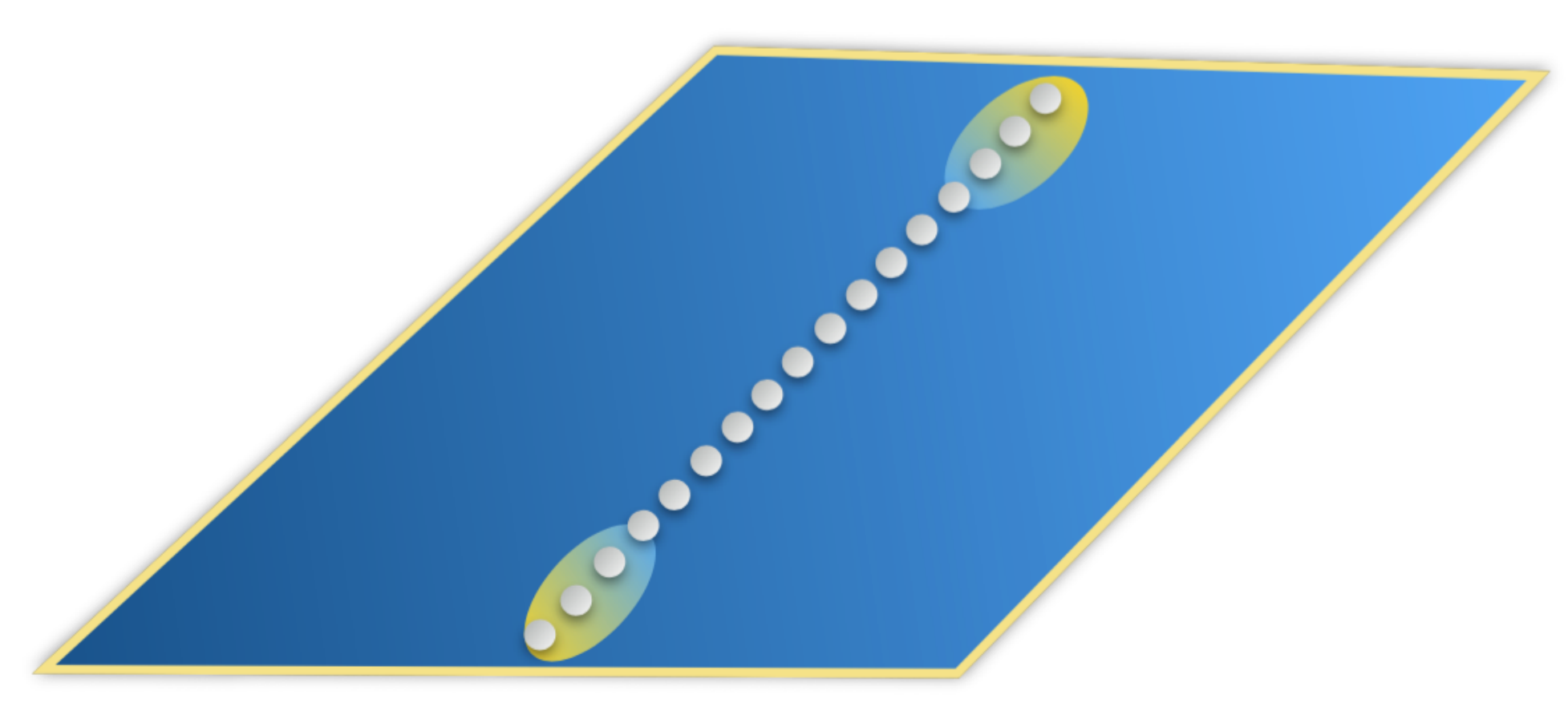}
\caption{One-dimensional wire formed by a chain of potential impurities fabricated on top of a $p$-wave superconductor. The topological nature of the wire is reflected in the Majorana end states.}\label{fig:schematic}
\end{figure}

\section{Model and the spectral problem}\label{model}

In this section, we study a problem of potential impurities on a spinless chiral $p$-wave superconductor. In Sec.~IV, we generalize our conclusions to the system including spin, and make use of the results obtained here.

\subsection{ Subgap spectrum of spinless $p$-wave model}

The studied system, depicted in Fig.~\ref{fig:schematic}, consists of electrons on a 2d superconductor with $N$ scalar impurities. This system is described by the Bogoliubov-de Gennes (BdG) Hamiltonian 
\begin{equation}\label{H_spinless}
\mathcal{H} = \xi_\vec{k}\tau_z +
\Delta (k_x\tau_x-k_y\tau_y) + U \sum_i \tau_z \delta(\vec{r}-\vec{r}_i) ,
\end{equation}
where $\xi_\vec{k} = k^2/(2m)-k_F^2/(2m)$ is the kinetic energy of the electrons with effective mass $m$ and Fermi wavenumber $k_F$, and the terms proportional to $\Delta$ represent the chiral $p$-pairing of electrons into Cooper pairs in the angular momentum channel $L_z=1$ . The matrices $\tau_x,\tau_y$ and $\tau_z$ correspond to Pauli matrices in particle-hole space throughout this text. The last term arises from the potential impurities with strength $U$ which could be attractive or repulsive. The Hamiltonian is expressed in the Nambu basis $\Psi(\vec r) = (\psi(\vec r),\psi^\dag(\vec r))^T$. The spectrum of the system can be solved from the BdG equation $\mathcal{H}\Psi=E\Psi$, which becomes
\begin{equation}
\begin{split}
\left[ E-\xi_\vec{k}\tau_z-\Delta(k_x\tau_x-k_y\tau_y) \right] \Psi(\vec r) \\
= U\sum_j \tau_z \delta(\vec{r}-\vec{r}_j) \Psi(\vec{r}_j) ,
\end{split}
\end{equation}
where we have isolated all impurity terms to one side.
Now we pass to momentum space using $\Psi(\vec{r}) = \int [d\vec{k}/(2\pi)^2] e^{i\vec{k} \cdot \vec{r}} \Psi_\vec{k}$, and obtain 
\begin{equation}
\begin{split}
\left[ E-\xi_\vec{k}\tau_z-\Delta(k_x\tau_x-k_y\tau_y) \right] \Psi_\vec{k} \\ = U \sum_j e^{-i \vec{k} \cdot \vec{r}_j} \tau_z \Psi(\vec{r}_j) .
\end{split}
\end{equation}
Multiplying by $\left[ E-\xi_\vec{k}\tau_z-\Delta(k_x\tau_x-k_y\tau_y) \right]^{-1}$ from the left and defining
\begin{equation}\label{J_def}
J_E(\vec{r}) = U \int \frac{d\vec{k}}{(2\pi)^2} \left[ E-\xi_\vec{k}\tau_z-\Delta(k_x\tau_x-k_y\tau_y) \right]^{-1} e^{i\vec{k}\cdot\vec{r}}
\end{equation}
we can go back to real space and evaluate the equation at $\vec{r}_i$ to obtain
\begin{equation}\label{BdG_separated}
\left[ 1- J_E(\vec{0})\tau_z \right] \Psi (\vec{r}_i) =  \sum_{j\neq i} J_E(\vec{r}_{ij}) \tau_z \Psi(\vec{r}_j)
\end{equation}
where $\vec{r}_{ij}=\vec{r}_i-\vec{r}_j$, and where we have separated the term $j=i$ from the sum. Thus we have re-formulated the original BdG eigenvalue problem as an eigenvalue problem for  $N$ coupled $2\times1$ spinors. 
We will restrict the chain of impurities to the $x$ direction by simply setting $y=0$ in the obtained expressions.
As seen in Appendix~\ref{NLEVP}, upon defining
$\gamma = 1 + \frac{\Delta^2}{v_F^2} $, $\beta = \Delta^2 k_F^2 - \gamma E^2$, and
$\tilde{\Delta} = \Delta^2\frac{k_F}{v_F \gamma}$, the expressions reduce to
\begin{align}
J_E(x\!=\!0) \approx& \frac{\alpha}{\sqrt{\beta}}
\left[ \tilde{\Delta} \tau_z - E \right] \label{J(r0)}\\
J_E(x\!\neq\! 0) \approx&\, \alpha \bigg[
\frac{-E}{\sqrt{\beta}} \RE\left(\Phi_0\right)
+ \Big( \frac{\tilde{\Delta}}{\sqrt{\beta}}\RE\left(\Phi_0\right) - \frac{1}{\gamma} \IM\left(\Phi_0\right) \Big) \tau_z \notag \\
- \frac{i\Delta \sign(x)}{\gamma}& \Big( \frac{1}{v_F}\Big[\frac{2}{\pi}-\RE\left(\Phi_1\right)\Big]
+ \frac{k_F}{\sqrt{\beta}} \IM\left(\Phi_1\right) \Big) \tau_x \bigg]	\label{J(r)}
\end{align}
where $\alpha=\pi\nu_0U$ characterizes the strength of the potential and 
$\Phi_n = I_n(x\Omega)-\textbf{L}_n(x\Omega)$, with 
$\Omega = \frac{1}{\gamma}\left(\frac{\sqrt{\beta}}{v_F} + ik_F\right) $. Here $I_n(x)$ and $\textbf{L}_n(x)$ are the modified Bessel and Struve functions of the first kind, and $v_F$ is the Fermi velocity. After inserting these expressions into Eq.~\eqref{BdG_separated} and writing it in a matrix form, we obtain
\begin{equation}\label{BdG_last_real}
\frac{1}{\sqrt{\beta}}\!
\begin{pmatrix}
(\varepsilon\!-1)A & B \\
B & (\varepsilon\!+1)A
\end{pmatrix}
\!\Psi = 
\begin{pmatrix}
C-\alpha^{-1} & D \\
D & \alpha^{-1}-C
\end{pmatrix}
\!\Psi,
\end{equation}
where $\varepsilon = E/\tilde{\Delta}$, $\Psi^T=[\Psi(x_1)^T\ldots \Psi(x_N)^T]$ is a $2N\times1$ spinor, and the $N\times N$ Hermitian submatrices $A,B,C,D$ are given in Appendix \ref{NLEVP}. The $2N\times 2N$ matrix structure appears due to the $N$ impurity sites each supporting two-component BdG spinors. This is a long-range tight binding problem expressed as a nonlinear eigenvalue problem, where the hopping elements between different impurity sites decay asymptotically as $\exp(-r/\xi_E)/\sqrt{k_Fr}$ for coherence length $\xi_E = \gamma v_F/\sqrt{\beta} $. We note that for low energies $\xi_E \approx \gamma v_F/(|\Delta| k_F) \equiv \gamma\xi$. The full solution to the matrix eigenvalue problem (\ref{BdG_last_real}) consists of $2N$ energy states with corresponding eigenspinors. The mathematical structure of the problem is closely analogous to those studied in Refs.~\onlinecite{weststrom:2015:1,poyhonen:2016:1,weststrom:2016:1} for magnetic chains. The nonlinear   energy dependence through $\beta$ and $\Phi_n$ complicates the solution considerably. 

\subsection{Solution to the nonlinear eigenvalue problem}

In this section, we will solve the eigenvalue problem given by Eq.~(\ref{BdG_last_real}). By considering a periodic lattice of impurities, we can make use of reciprocal space and make analytical progress. To this end, we define Fourier transforms of the submatrices, given by 
\begin{equation}
a_k = \sum_j A_{ij}e^{ika(i-j)},
\end{equation}
with analogous expressions for $b_k, c_k,$ and $d_k$.

The earlier separation of energy dependence into factors outside the submatrices makes the transformation into reciprocal space straightforward; the resultant equation is practically identical to Eq.~\eqref{BdG_last_real}:
\begin{equation}\label{BdG_k}
\frac{1}{\sqrt{\beta}}
\!\begin{pmatrix}
(\varepsilon\!-\!1)a_k & b_k \\
b_k & (\varepsilon\!+\!1)a_k
\end{pmatrix}\!
\psi_k\! =\! 
\begin{pmatrix}
c_k\!-\!\alpha^{-1} & d_k \\
d_k &  \alpha^{-1}\!-c_k
\end{pmatrix}\!
\psi_k
\end{equation}
with the submatrices replaced by their Fourier transforms as defined above. It is, however, now possible to re-express the eigenvalue equation in a more tractable form. By moving the right-hand-side matrix to the left-hand side, we obtain an equation of the form $G^{-1}\psi_k=0$, with
\begin{equation}\label{G_def}
G^{-1}\! =\! 
\!
\!\begin{pmatrix}
(\varepsilon\!-\!1)a_k -\sqrt{\beta}(c_k\!-\!\frac{1}{\alpha})& b_k - \sqrt{\beta}d_k \\
b_k -\sqrt{\beta}d_k & (\varepsilon\!+\!1)a_k + \sqrt{\beta}(c_k\!-\!\frac{1}{\alpha})
\end{pmatrix}
\end{equation}
As detailed in Appendix~\ref{SUBGAP}, the energy bands can be solved from the condition $\mathrm{det}G^{-1}=0$. This provides a closed form equation for the energy bands $E_k$ in the form
\begin{equation}\label{eq_E^4}
P_{2,k}\beta(E_k) + P_{1,k}\sqrt{\beta(E_k)} + P_{0,k} = 0,
\end{equation}
where the coefficients $P_i$ are given in the appendix. In addition to the explicit $E_k$ terms in Eq.~\eqref{eq_E^4}, there is an additional energy dependence through $P_i$ hidden in the notation, which ultimately originates from $\Phi_n(x\Omega)$ in Eq.~\eqref{J(r)}. Fortunately, this energy dependence turns out to be negligible for the energy bands, which can be solved by treating  Eq.~(\ref{eq_E^4}) as a polynomial for $E_k$ with $P_i \to P_i(E = 0)$. As expected from particle-hole symmetry, the two solutions for each $k$ come in pairs $\pm E_k$ which in Eq.~\eqref{eq_E^4} follows from the fact that $\beta$ is a function of the square of $E_k$.

\begin{figure}
\includegraphics[width=0.95\linewidth]{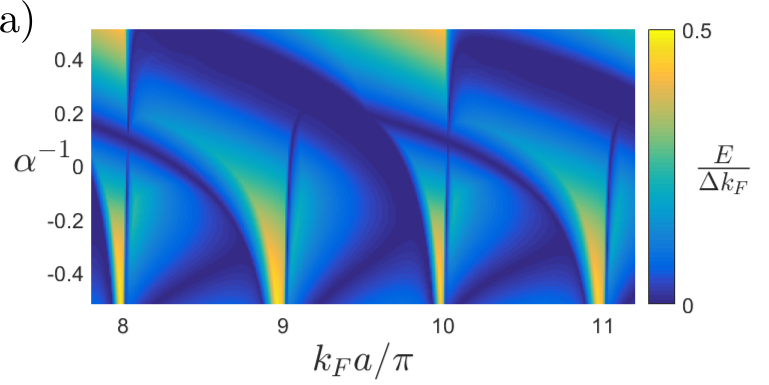}
\includegraphics[width=0.95\linewidth]{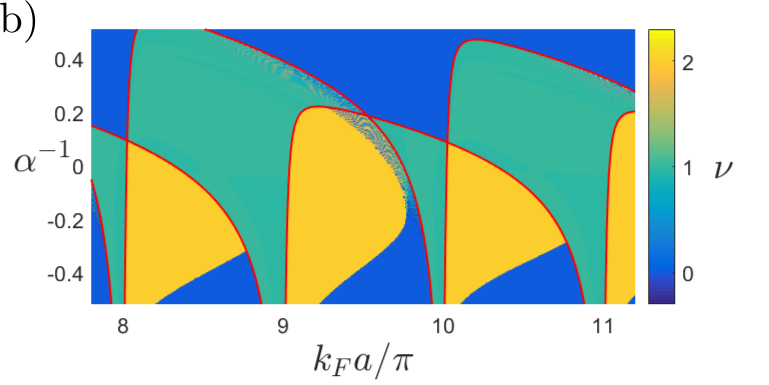}
\includegraphics[width=0.95\linewidth]{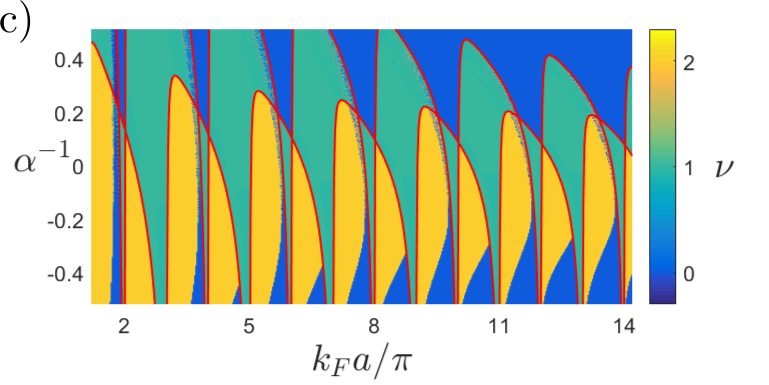}

\caption{
(a) Lowest positive energy eigenvalue as a function of $k_Fa$ and $\alpha^{-1}$.
The values used for the coherence length is $\xi = 20a$, with $\Delta/v_F = 1/(k_F\xi)$. For this plot, we have set $v_F = 100$.
(b) Winding number as a function of $k_Fa$ and $\alpha^{-1}$, over the same parameter range as in (a). The red curve corresponds to the analytical solution in $k$-space for boundaries between different parity phases, calculated using Eq.~\eqref{alpha_parity}. 
(c) Same as in (b), but with a wider range of $k_Fa$ values.
}
\label{top_diagram}
\end{figure}

In Fig.~\ref{top_diagram}(a) we have plotted the minimum of the positive energy solution $\min_k E_k$ as a function of the dimensionless impurity strength $\alpha=\pi\nu_0U$ and $k_Fa$, where $a$ is the lattice constant of the impurity lattice. This shows that the 1d system is generally gapped and that there exist distinct regions separated by gap closing transitions. One peculiar feature of the spectrum is that in the limit $\alpha^{-1}\to 0^{\pm}$, marking an infinite repulsion or attraction depending on the sign, the spectrum is continuous. This counterintuitive feature follows from the properties of the single-impurity bound states which coincide for  infinite repulsion and attraction.\cite{kaladzhyan:2016:2,kim:2015:1}  

\section{Topological properties}\label{topo}
When studying the topological properties of a system, it is convenient to use the Hamiltonian for the system. Since we have formulated the problem in terms of a nonlinear eigenvalue probem, it is not obvious what our Hamiltonian should be. However,
following Refs.~\onlinecite{poyhonen:2016:1,weststrom:2016:1}, the topological phase diagram of an impurity chain can be extracted by identifying an effective Hamiltonian possessing the same topological properties as our system.
This Hamiltonian can be obtained from Eq.~\eqref{G_def} by defining $\tilde H_k \equiv G^{-1}(0)$. This immediately gives
\begin{equation}\label{H_eff}
\tilde{H}_k =
\left[  
	\tilde a_k +\left|\Delta\right| k_F( \tilde c_k - \frac{1}{\alpha} )\right]\tau_z + \left[\left|\Delta\right| k_F\tilde d_k - 
	\tilde b_k\right]\tau_x.
\end{equation}
Here we have defined 
$ \tilde a_k = \lim_{E\to 0} a_k $
for all $a_k,...,d_k$.
The effective Hamiltonian in Eq.~(\ref{H_eff}) has a few key properties that make it useful for our purposes. Due to the above construction, the gap closings of the effective Hamiltonian coincide exactly with those of the parent model, and can easily be calculated from Eq.~\eqref{H_eff}. Additionally, in the limit of an infinite system, any zero-energy edge modes in the original system will be reproduced exactly in the effective model. The two systems are thus topologically equivalent. So while $\tilde H_k$ does not yield the correct energy spectrum or states of the system, it is still useful in studying all its relevant topological features.

\subsection{Winding number and $\mathbb{Z}_2$ phases }

To extract the topological phase diagram of the 1d chain, we need to identify the relevant symmetries of the model. Since we are studying a superconductor, the particle-hole symmetry that is built in the BdG formalism is naturally present. This symmetry puts the system in the Altland-Zirnbauer class D which supports a $\mathbb{Z}_2$ classification,\cite{schnyder:2009:1,ryu:2010:1} distinguishing phases with different fermion parity.\cite{kitaev:2001:1}  However, a reduction of the original 2d model (\ref{H_spinless}) into an effectively 1d model gives rise to additional chiral symmetry. Chiral symmetry manifests as the anticommutation property $\lbrace \tilde{H},\mathcal{C} \rbrace = 0$ with the chiral symmetry operator $\mathcal{C} = \tau_y$. With the additional symmetry, the 1d model belongs to the BDI class, supporting a $\mathbb{Z}$-valued winding number invariant.\cite{schnyder:2009:1,ryu:2010:1} We adopt the BDI classification below, but also plot the $\mathbb{Z}_2$ phase boundaries which serve as an additional consistency check for the results.  

The topological winding number of a 1d system\cite{volovik} can be obtained through the formula
\begin{equation}\label{wind_nr}
\nu = \frac{i}{4\pi} \int_{-\pi/a}^{\pi/a}
\mathrm{d}k ~\text{tr} 
\left[ \mathcal{C} {H}^{-1} \partial_k {H} \right].
\end{equation}
As discussed earlier in this work, we can obtain the correct result by simply inserting the effective Hamiltonian $\tilde H$ in the above equation.
In Fig.~\ref{top_diagram} we have plotted the winding number \eqref{wind_nr} as a function of the relevant system parameters. We find values
$\nu=0,1,2$ describing three topologically distinct phases. In accordance with general principles, the topological phase transitions can occur only when the energy gap of the systems closes. Below, we will verify that the winding number is directly related to the number of the Majorana end states in finite wires. 

The $\mathbb{Z}_2$ invariant, measuring the fermion parity, can change its value only when the energy gap closes at the points $k=0$ and $k=\pi/a$. The $\mathbb{Z}_2$ phase boundaries can be calculated from the gap-closing condition of the effective Hamiltonian \eqref{H_eff}. Thus, keeping other parameters constant, the gap closings can be parametrized as a curve $(k_Fa,\alpha)$. Due to the antisymmetric nature of the submatrices $B,D$, the Fourier transformed coefficients $\tilde b_k, \tilde d_k$ vanish identically at $k=0,\pi/a$, and the phase boundary condition becomes 
\begin{equation}\label{alpha_parity}
\alpha^{-1} = \bigg( \frac{\tilde a_k}{\left|\Delta\right| k_F}  + \tilde c_k \bigg) \Bigg\rvert_{k=0,\pi/a}
\end{equation}
which are plotted in Figs.~\ref{top_diagram} (b) and (c) together with the $\mathbb{Z}$-valued invariant. The graphs provide an independent way of establishing the phase boundaries between phases of different parity. 

\subsection{Majorana bound states }
\begin{figure}
\includegraphics[width=0.47\columnwidth]{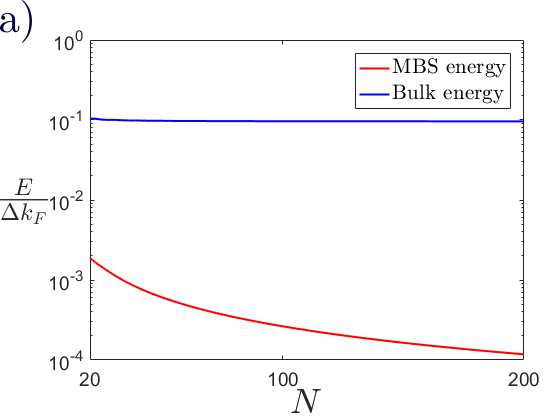}
\includegraphics[width=0.47\columnwidth]{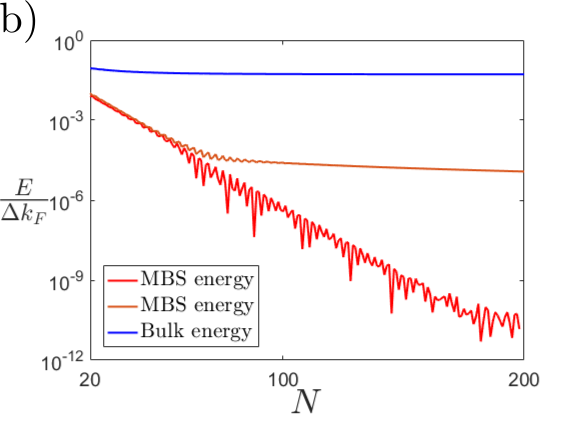}
\includegraphics[width=0.47\columnwidth]{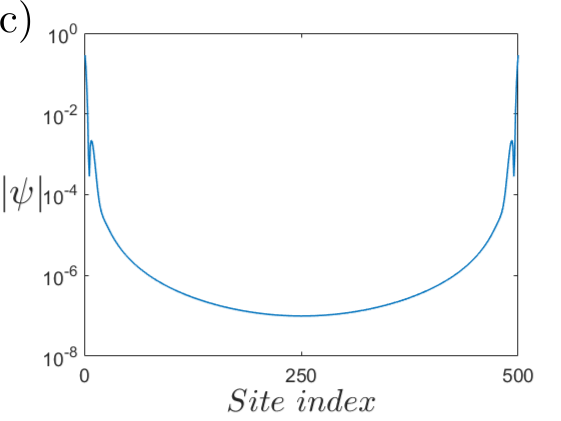}
\includegraphics[width=0.47\columnwidth]{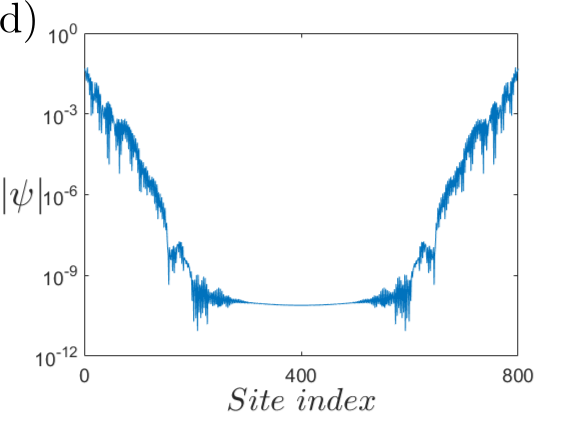}
\caption{a) Dependence of the bulk and MBS energy in the $\nu=1$ phase on the length $N$ of the impurity chain in blue and red, respectively.
b) Same as in a), except for the $\nu=2$ phase.
c) Localization of the MBS wavefunction in the $\nu=1$ phase, with chain length $N=500$.
d) Same as in c), except for the $\nu=2$ phase, and with chain length $N=800$. \\
Parameters used in the calculations above are $\xi = 20a$, $k_Fa = 8.5 \pi$; a) and c) have $\alpha^{-1}=0.2$, while b) and d) have $\alpha^{-1}=-0.2$. In all cases, $\Delta/v_F = 1/(k_F\xi)$.}\label{fig:majoranas}
\end{figure}

The observable difference between the distinct topological phases becomes apparent in finite systems. The physical significance of the value
$\left| \nu \right|$ is the number of zero-energy Majorana bound states localized at each end of the impurity chain. The $\nu=0$ phase corresponds to the topologically trivial phase with even-parity ground-state and supports no end states. In Figs.~\ref{fig:majoranas} (a) and (b), we have plotted the lowest few positive-energy eigenvalues of a finite system as a function of chain length in the $\nu = 1$ and $\nu = 2$, respectively. As can be seen in the figure, as the chain length increases, the expected number of low-energy states separate from the bulk gap and go to lower energies. We interpret these as MBS at the edges of the chain. In Figs.~\ref{fig:majoranas} (c) and (d) we have plotted the amplitude of the corresponding MBS eigenstates at chain lengths 500 and 800, respectively. As seen in the figure, the states are localized at the edges of the chain. We have omitted the second MBS in (d), as it looks similar to the one already displayed.

\section{Including spin}\label{spin}

So far we have considered impurities on a spinless chiral $p$-wave superconductor. In this section, we generalize the results for representative models with spin. One way to parametrize the different 2d $p$-wave models with spin is to write the BdG Hamiltonian of the underlying SC in the Nambu basis $\Psi = (\psi_\uparrow,\psi_\downarrow,\psi_\downarrow^\dag,-\psi_\uparrow^\dag)^T$ as
\begin{equation}
\mc H = \begin{pmatrix}
\xi_\vec{k}  & \Delta\vec d \cdot \bm \sigma\\
\Delta(\vec d \cdot \bm \sigma)^\dagger & -\xi_\vec{k} 
\end{pmatrix}.
\end{equation}
Here the $\vec{d}$-vector determines the spin-structure of the Cooper pairs, and $\bm \sigma$ is a vector of Pauli matrices in spin space. In this work, we will concentrate on two cases for the $\vec{d}$-vector, which contains the information about the superconducting triplet pairing.

\subsection{Chiral superconductor}
The out-of-plane pairing with $\vec{d}=(0,0,k_x+ik_y)$ is probably the experimentally most relevant case at the moment, since it is the main candidate to describe pairing in Sr$_2$RuO$_4$. The resulting Hamiltonian
\begin{equation}\label{Ha}
\mathcal{H}_\text{a} = \xi_\vec{k}\tau_z +
\Delta \sigma_z(k_x\tau_x-k_y\tau_y) + U \sum_i \tau_z \delta(\vec{r}-\vec{r}_i)
\end{equation}
corresponds to the pairing of antiparallel spins in the $L_z=1$ channel, leading to a ground state with spontaneous broken time-reversal symmetry. In this form we see that the matrix structure is that of two spinless Hamiltonian blocks with different signs of $\Delta$.

\begin{figure}
\includegraphics[width=0.95\linewidth]{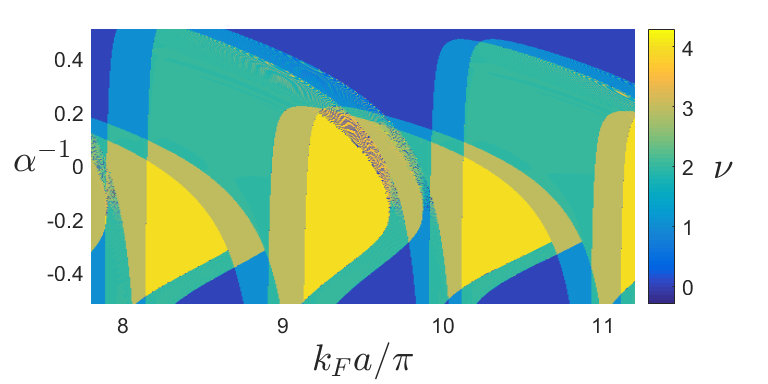}
\caption{
The winding number for both the chiral and helical superconductor with a Zeeman field of magnitude $B$. In the chiral superconductor, it is parallel to the surface, while it is normal to the surface in the helical superconductor. Here $2|B|m = 8\cdot 10^3/\xi^2$ with all other parameters being the same as in Fig.~\ref{top_diagram}.}
\label{fig:top_diag_b}
\end{figure}

Since the Hamiltonian $\mathcal{H}_\text{a}$ is diagonal in spin space, the results for the spinless system have straightforward generalizations. Indeed, the calculations for the spinful case proceed largely identically, yielding an effective Hamiltonian
\begin{equation}\label{Ha_eff}
\tilde H_{\text{a},k} = \left[ \tilde a_k + \tilde c_k - \frac{1}{\alpha} \right]\tau_z\sigma_z + \left[\tilde d_k - 
	\frac{1}{\left|\Delta\right| k_F} 
	\tilde b_k\right]\tau_x.
\end{equation}
which can easily be seen by noting that the left term is symmetric under the change $\Delta \to -\Delta$, whereas the right term is antisymmetric. Consequently we can define the chiral symmetry operator as
$\mathcal{C} = \sigma_z \tau_y $, and the winding number Eq.~\eqref{wind_nr} can immediately be seen as the sum of the two blocks of spinless systems, resulting in a change $ \nu \to 2 \nu$. Hence, the topological phase diagram of an impurity chain on chiral spinful system coincides with the spinless case with the winding numbers doubled. With an additional interaction that lifts the spin-degeneracy without destroying chiral symmetry, it is also possible to access the odd winding number phases $ \nu=1,3$. This could be accomplished by, for example, adding an in-plane Zeeman field $B\sigma_x$ to Eq. ~\eqref{Ha}. The unitary transformation $\mathcal H_a\to U\mathcal H_a U^\dagger$, with $U = \exp(i\frac{\pi}{4}\tau_z\sigma_y)$, only acts nontrivially on the new magnetic field term, and results in a transformed Hamiltonian
\begin{equation}\label{Ha_B}
\mathcal{H}_\text{a} = (\xi_\vec{k} + B\sigma_z)\tau_z +
\Delta \sigma_z(k_x\tau_x-k_y\tau_y) + U \sum_i \tau_z \delta(\vec{r}-\vec{r}_i).
\end{equation}
Hence in each block, the chemical potential is shifted by an amount $\pm B$, or -- equivalently -- there is a shift $k_F \to \sqrt{k_F^2 \pm 2mB}$. While the invariants of the two blocks are still to be simply added together, their topological diagrams are now shifted in parameter space with respect to each other, resulting in the appearance of new, odd-$\nu$ phases. This behavior can be seen in Fig.~\ref{fig:top_diag_b}. We note that the choice of $x$ direction on the magnetic field here was arbitrary, and it can equally be in any in-plane direction; an in-plane magnetic field with angle $\phi$ to the $x$ axis can be rotated back to the axis by application of the unitary transformation $\exp(i\frac{\phi}{2}\sigma_z)$, which again commutes with the remainder of the Hamiltonian.

\subsection{Helical superconductor}
Another important $p$-wave parent state for the impurity chain is realized in the case of an in-plane $\vec{d}$-vector of the form $\vec{d} = (k_x,k_y,0)$. The Hamiltonian for this model is
\begin{equation}
\mathcal{H}_\text{p} = \xi_\vec{k} \tau_z + \Delta \tau_x (k_x \sigma_x + k_y \sigma_y) + U \sum_i \tau_z \delta(\vec{r}-\vec{r}_i)
\end{equation}
and describes a time-reversal invariant system $T\mathcal{H}_\text{p}(\vec{k})T^{-1}=\mathcal{H}_\text{p}^*(\vec{-k})$, where the time-reversal operator is  $T=i\sigma_y\mathcal{K}$ with $\mathcal{K}$ acting as a complex conjugation. The spin up and spin down electrons condense in the opposite angular momentum channels $L_z=\pm1$. This Hamiltonian describes a $\mathbb{Z}_2$  topological superconductor with helical edge states.\cite{bernevig} 

With a change of the basis $ (\psi_\uparrow,\psi_\downarrow,\psi_\downarrow^\dag,-\psi_\uparrow^\dag)^T\to (\psi_\uparrow,\psi_\uparrow^\dag,\psi_\downarrow,\psi_\downarrow^\dag)^T$, the in-plane Hamiltonian takes the form
\begin{equation}\label{Hp}
\mathcal{H}_\text{p} = \xi_\vec{k}\tau_z -\Delta \sigma_z(k_x\tau_x+k_y\sigma_z\tau_y) + U \sum_i \tau_z \delta(\vec{r}-\vec{r}_i)
\end{equation}
Thus, in the new basis $\mathcal{H}_\text{p}$ differs from the chiral case \eqref{Ha} only by the sign of $\Delta$ and the $\sigma_z$ matrix multiplying the $k_y$-term. The sign of 
$\Delta$ is mostly immaterial, so the relevant difference arises from the $k_y$-term in Eq.~\eqref{Hp}. However, in deriving of the effective 1d model, the difference arising from the $k_y$-term vanishes when the impurity chain is formed in the $x$ direction. Therefore, the effective 1d model is the same as for the chiral case with the replacement $\Delta\to -\Delta$. For the same reasons as discussed above, the energy gap and the topological phase diagram for the impurity chain on a helical superconductor coincides with the spinless case, with the winding number doubled $\nu \to 2 \nu$. Thus, the wire supports three topologically distinct phases. Analogously to the chiral case, odd-$\nu$ phases can be accessed by applying a magnetic field to the system. However, for the helical superconductor, this added field must be out-of-plane to preserve the requisite symmetries. Under the change of basis detailed above, the magnetic field term $B\sigma_z$ takes the form $B\tau_z\sigma_z$ automatically, and hence no additional unitary transformations are necessary. The main difference to the chiral case is then the requirement that the magnetic field be locked to the $z$ axis rather than being allowed to rotate in a plane.

\section{Discussion and summary} \label{summ}
In this work we have studied the topological and spectral properties of 1d chains formed by scalar impurities deposited on 2d $p$-wave superconductors. The main contribution of the paper is the formulation of the theoretical framework which allows the solution of the spectrum and the topological phase diagram of the 1d system. Our approach enables a calculation of the topological energy gaps even when they are significant compared to the substrate pairing gap. In finite systems, the topological gap separates the Majorana end states from the bulk states and is directly relevant to their experimental observation.  We found that in ideal circumstances the topological gap may be a significant fraction of the pairing gap of the substrate. This makes the direct observation of the midgap Majorana states in 1d systems significantly more favourable compared to the 2d vortex core Majorana states that are separated from the other quasiparticle states by a minigap $(\Delta k_F)^2/E_F$ (where $E_F = \frac{1}{2m}k_F^2$ is the Fermi energy), which is very small in a weak-coupling superconductor. 

We found that 1d impurity wires on chiral and helical 2d $p$-wave superconductors may display at least three different topological phases. The fabrication of these 1d chains and the observation of the topological end states could be carried out by scanning tunneling microscopy methods that have been used in studying  magnetic chains.\cite{nadj-perge:2014:1} Since the topological phase depends on the lattice constant of the impurity chain, the different phases could be accessed by fabricating chains with different lattice constants. An abrupt change of the lattice constant in the middle of a chain could realize a topological phase boundary with localized domain-wall states. 

The effective low-energy theory of a scalar impurity chain is quite analogous to that of a magnetic chain, so the role of disorder is expected to display similar features. In this picture, missing impurity sites give rise to vacancy states below the topological band edge and dilute vacancy states eventually drive the system gapless.\cite{weststrom:2016:1} However, a dilute concentration of defects do not destroy the topological phases.     

The most promising candidate for a chiral $p$-wave superconductor at the moment is Sr$_2$RuO$_4$ which could be employed as a substrate.
Artificial $p$-dominated chiral superconductors could also be expected to exhibit qualitatively similar physics. Although we considered a subset of all possible $p$-wave superconductors, it is plausible that the nontrivial phases in the 1d chain arise as a consequence of the nontrivial 2d topology. One can also regard the study of topology of a 1d chain as a diagnostic tool to study the nature of the substrate.

\acknowledgments
The authors acknowledge the Academy of Finland and the Aalto Center for Quantum Engineering for support. KP acknowledges the Finnish Cultural Foundation for support.

\appendix
\numberwithin{equation}{section}
\section{Spectrum as a nonlinear eigenvalue problem}\label{NLEVP}

Here we present how to bring the Bogoliubov-de Gennes equation \eqref{BdG_separated} into the form of a nonlinear $2N \times 2N$  matrix eigenvalue equation. 
The equation at hand is
\begin{equation}\label{BdG_sep_app}
\left[ 1- J_E(\vec{0})\tau_z \right] \Psi (\vec{r}_i) =  \sum_{j\neq i} J_E(\vec{r}_{ij}) \tau_z \Psi(\vec{r}_j) ,
\end{equation}
where 
\begin{equation}
J_E(\vec{r}) = U \int \frac{d\vec{k}}{(2\pi)^2}
\left[ E-\xi_\vec{k}\tau_z-\Delta(k_x\tau_x-k_y\tau_y) \right]^{-1} e^{i\vec{k}\cdot\vec{r}} .
\end{equation}
Inverting the matrix in the integrand, we can write the functions $J_E$ in the form
\begin{equation}
J_E(\vec{r}) = U \int \frac{d\vec{k}}{(2\pi)^2}
\frac{E+\xi_\vec{k}\tau_z+\Delta(k_x\tau_x-k_y\tau_y)}{E^2-\xi_\vec{k}^2-\Delta^2 k^2} e^{i\vec{k}\cdot\vec{r}}
\end{equation}
The integral diverges at $r = 0$, ultimately because of the limitations of the BCS model as a low-energy theory. We will introduce an artificial cutoff by assuming $k$ is close to the Fermi level so that $k\approx k_F + \xi/v_F$, as is standard in the field. This will be done in the integrals throughout this section.

For $\vec{r}=0$, the terms including $k_x$ and $k_y$ vanish under angular integration, and the remaining terms are independent of the angle, so we obtain, after linearization,
\begin{equation}
J_E(\vec{0}) = U\nu_0 \int_{-\infty}^\infty d\xi_\vec k
\frac{E + \xi_\vec{k}\tau_z}{E^2-\xi_\vec{k}^2-\Delta^2 (\xi_\vec k/v_F + k_F)^2}.
\end{equation}
This form lends itself to use of the residue formula, yielding
\begin{equation}
J_E(\vec{0}) = \frac{\alpha}{\sqrt{\beta}}
\left[ \tilde{\Delta} \tau_z - E \right]
\end{equation}
where $\alpha = \pi\nu_0 U$, $\beta = \Delta^2 k_F^2 - \gamma E^2$ and $\tilde{\Delta} = \Delta^2\frac{k_F}{v_F \gamma}$ with $\gamma = 1 + \frac{\Delta^2}{v_F^2}$.

For nonzero $\vec{r}$, the angular integral gives two Bessel functions of the first kind:
\begin{align}
& \int d\vec{k}
\frac{E+\xi_\vec{k}\tau_z+\Delta(k_x\tau_x-k_y\tau_y)}{E^2-\xi_\vec{k}^2-\Delta^2 k^2} e^{i\vec{k}\cdot\vec{r}} = \notag \\
2\pi &\int_0^\infty  dk k
\frac{\left(E + \xi_\vec{k}\tau_z \right)J_0(kr) + ik \frac{\Delta}{r} \left( x\tau_x-y\tau_y \right)J_1(kr)}{E^2-\xi_\vec{k}^2-\Delta^2 k^2}
\end{align}
After linearization, use of a suitable representation for the Bessel functions brings the remaining integrals into a form that can readily be solved through residue integration, leading to
\begin{align}
J_E(x\!\neq\! 0) =&\, \alpha \bigg[
\frac{-E}{\sqrt{\beta}} \RE\left(\Phi_0\right)
+ \Big( \frac{\tilde{\Delta}}{\sqrt{\beta}} - \frac{1}{\gamma} \IM\left(\Phi_0\right) \Big) \tau_z \notag \\
- \frac{i\Delta \sign(x)}{\gamma}& \Big( \frac{1}{v_F}\Big[\frac{2}{\pi}-\RE\left(\Phi_1\right)\Big]
+ \frac{k_F}{\sqrt{\beta}} \IM\left(\Phi_1\right) \Big) \tau_x \bigg]
\end{align}
where we have set $y$ to zero due to the one-dimensional structure of the system, as in the main text. We have here defined the function
$\Phi_n \equiv I_n(x\Omega)-\textbf{L}_n(x\Omega)$,
where
$\Omega = \frac{1}{\gamma}\left(\frac{\sqrt{\beta}}{v_F} + ik_F\right) $, and $I_n(x)$ and $\textbf{L}_n(x)$ are the modified Bessel and Struve functions of the first kind, respectively.

After inserting the obtained expressions into the BdG equation \eqref{BdG_separated}, we collect terms with the same prefactors and define the submatrices
\begin{align}\label{real_matrices}
A_{ij} &= \tilde \Delta[\delta_{ij}
+ \left(1-\delta_{ij}\right) \RE\Phi_{ij}^0] \\
B_{ij} &= i \left(\delta_{ij}-1\right) \gamma^{-1} \Delta k_F(\IM\Phi_{ij}^1) \frac{x}{r} \\
C_{ij} &= \left(\delta_{ij}-1\right) \gamma^{-1} \IM\Phi_{ij}^0 \\
D_{ij} &= i \left(1-\delta_{ij}\right) \frac{\tilde{\Delta}}{\Delta k_F} \left(\tfrac{2}{\pi}-\RE\Phi_{ij}^1\right)
	\frac{x}{r} 
\end{align}
which are clearly Hermitian. Here we use the shorthand
$x \equiv x_{ij} \equiv x_i-x_j$ and
$r \equiv \left|x_{ij}\right|$,
and $\Phi_{ij}^n \equiv \Phi_n(x_{ij})$ for the special functions.
We can thus bring the nonlinear eigenvalue equation into the form \eqref{BdG_last_real}.

\section{Solution to the subgap spectrum}\label{SUBGAP}
In this appendix, we derive an equation for the eigenvalues of Eq.~\eqref{BdG_k}. We start out with $G^{-1}(E_k)\psi_k=0$, where
\begin{equation}\label{M_def_app}
G^{-1}\! =\! 
\!
\!\begin{pmatrix}
(\varepsilon\!-\!1)a_k -\sqrt{\beta}(c_k\!-\!\frac{1}{\alpha})& b_k - \sqrt{\beta}d_k \\
b_k -\sqrt{\beta}d_k & (\varepsilon\!+\!1)a_k + \sqrt{\beta}(c_k\!-\!\frac{1}{\alpha})
\end{pmatrix}.
\end{equation}
As noted in the main text, the energy dependence in the Fourier transformed matrix elements has a negligible impact on the solution of the nonlinear eigenvalue problem. Hence we can set the energy in them to zero, which in the main text is denoted by
$a_k \to \tilde a_k$, but here we will forego that notation with no risk of confusion. By calculating the determinant, we observe that all remaining energy-dependence outside of $\beta$ is a single term proportional to $\varepsilon^2$. Since we can write this in terms of $\beta = \Delta^2 k_F^2 - \gamma \tilde{\Delta}^2\varepsilon^2$, finding the zeros for the determinant reduces to solving a quadratic equation in $\sqrt{\beta}$
\begin{equation}
P_2 \beta + P_1 \sqrt{\beta} + P_0 = 0
\end{equation}
where the coefficients $P_i$ are given by the expressions 
\begin{align*}
P_2 &= \frac{a^2}{\gamma\tilde{\Delta}^2} + c^{\prime 2} + d^2 \\
P_1 &= 2(ac^\prime - bd) \\
P_0 &= a^2(1-\gamma\frac{v_F^2}{\Delta^2}) + b^2.\\
\end{align*}
The quadratic equation has two solutions, but it turns out that for all relevant parameters, one of the solutions is negative, which is not sensible for $\sqrt{\beta}$. The energy is then given by the positive solution for $\sqrt{\beta}$ through the relation
\begin{equation}
	|E_k| = \sqrt{(\Delta^2k_F^2 - \beta_k)\gamma^{-1}}.
\end{equation}


\end{document}